\documentclass{aa}  

\usepackage{graphicx}
\usepackage{txfonts}
\usepackage{amsmath}
\usepackage{amssymb}
\usepackage{natbib}

\graphicspath{{./figures/}}
\bibpunct{(}{)}{;}{a}{}{,}
\newlength{\minuslength}
\begin{document}

   \title{ALMA observations of water deuteration: A physical diagnostic of the formation of protostars}


   \author{S. S. Jensen\inst{1}\thanks{\email{sigurd.jensen@nbi.ku.dk}}
             \and J. K. J{\o}rgensen\inst{1}
          \and L. E. Kristensen\inst{1}
          \and K. Furuya\inst{2}        
          \and A. Coutens\inst{3}
          \and E. F. van Dishoeck\inst{4,5}
          \and D. Harsono\inst{4}
          \and M. V. Persson\inst{6}}

   \institute{Niels Bohr Institute \& Centre for Star and Planet Formation, University of Copenhagen, {\O}ster Voldgade 5-7, DK-1350 Copenhagen K, Denmark
	\and Center for Computational Sciences, University of Tsukuba, Japan
	\and Laboratoire d'Astrophysique de Bordeaux, Univ. Bordeaux, CNRS, B18N, all{\'e}e Geoffroy Saint-Hilaire, 33615 Pessac, France
	\and Leiden Observatory, Leiden University, PO Box 9513, 2300 RA Leiden, The Netherlands
	\and Max--Planck Institute f{\"u}r extraterrestrische Physik (MPE), Giessenbachstrasse, 85748 Garching, Germany
	\and Department of Space, Earth and Environment, Chalmers University of Technology, Onsala Space Observatory, 439 92 Onsala, Sweden}

   \date{Draft date: \today}

 
  \abstract
   {How water is delivered to planetary systems is a central question in astrochemistry. The deuterium fractionation of water can serve as a tracer for the chemical and physical evolution of water during star formation and can constrain the origin of water in Solar System bodies.}
   {The aim is to determine the HDO/H$_2$O ratio in the inner warm gas toward three low-mass Class 0 protostars  selected to be in isolated cores, i.e., not associated with any cloud complexes. Previous sources for which the HDO/H$_2$O ratio have been established were all part of larger star-forming complexes. Determining the HDO/H$_2$O ratio toward three isolated protostars allows comparison of the water chemistry in isolated and clustered regions to determine the influence of local cloud environment.}
   {We present ALMA Band 6 observations of the HDO $3_{1,2}$--$2_{2,1}$ and $2_{1,1}$--$2_{1,2}$ transitions at 225.897~GHz and 241.562~GHz along with the first ALMA Band 5 observations of the H$_2^{18}$O $3_{1,3}$--$2_{2,0}$ transition at 203.407~GHz. The high angular resolution observations (0\farcs3-1\farcs3) allow the study of the inner warm envelope gas. Model-independent estimates for the HDO/H$_2$O ratios are obtained and compared with previous determinations of the HDO/H$_2$O ratio in the warm gas toward low-mass protostars.}
   {We successfully detect the targeted water transitions toward the three sources with S/N > 5.
   We determine the HDO/H$_2$O ratio toward L483, B335 and BHR71--IRS1 to be ($2.2\pm0.4$)$\times 10^{-3}$, ($1.7\pm0.3$)$\times 10^{-3}$, and ($1.8\pm0.4$)$\times 10^{-3}$, respectively, assuming $T_\mathrm{ex} = 124$~K. The degree of water deuteration of these isolated protostars are a factor of 2--4 higher relative to Class 0 protostars that are members of known nearby clustered star-forming regions.}
   {The results indicate that the water deuterium fractionation is influenced by the local cloud environment. This effect can be explained by variations in either collapse timescales or temperatures, which depends on local cloud dynamics and could provide a new method to decipher the history of young stars.}


   \maketitle
%
\section{Introduction}
How water evolves during star formation, from the molecular cloud core down to the protoplanetary disk, is a key question concerning the origin of the Solar System, the formation of exoplanets, and, ultimately, the possible emergence of life in planetary systems. 

Water is observed during all stages of star formation: from molecular clouds, through the dense-core phase and the protoplanetary disk and finally in planetary systems such as our Solar System \citep{dishoeck2014}. In star-forming pre-stellar molecular clouds, water is predominantly present in the ice: the gas-phase abundance is constrained to be of the order $X_\mathrm{gas}$(H$_2$O) $\sim10^{-8}$--$10^{-9}$ relative to H$_2$ \citep{bergin2002, caselli2012} while absorption ice spectroscopy has revealed $X_\mathrm{ice}$(H$_2$O) $\sim10^{-4}$--$10^{-5}$ \citep{pontoppidan2004}. It is an open question what processing, if any, water experiences during star formation. Classically, two scenarios are considered: inheritance or processing. In the inheritance scenario, the bulk of the water present in the planetary system after formation is inherited from the molecular cloud and is therefore representative of the chemistry in the molecular cloud before star formation begins \citep[e.g.,][]{visser2009, drozdovskaya2016}. In the alternative scenario, a substantial amount of water is destroyed and reformed in local processes within the envelope or protoplanetary disk during the formation process. In this case the final water chemistry is determined by local processes in the specific system \citep[see, e.g.,][ for an discussion of each scenario]{cleeves2014}.
 
The deuterium fractionation of water is a useful proxy for the processing of water during star formation and can help distinguish between the two scenarios outlined above.
The enrichment of deuterium is driven by several chemical processes, and their efficiency depend on physical conditions such as temperature, density, visual extinction, and ionization sources \citep{ceccarelli2014}. 
At low temperatures the prominent pathway is the gas-phase exchange reaction H$_{3}^{+}$ + HD $\rightleftharpoons$ H$_2$D$^{+}$ + H$_2$ + $\Delta E$, where $\Delta E \approx124$~K\footnote{The exact value of $\Delta E$ depends on the spin state of the involved reactants.}. This reaction is effectively one-way at low temperatures ($T \lesssim 50$~K) due to the endothermicity of the backward reaction. This leads to an enrichment of H$_2$D$^+$ which subsequently dissociatively recombines with free electrons to form atomic D, thus increasing the local atomic D/H ratio in the gas-phase and ultimately on dust grain surfaces where water and other molecules are formed through hydrogenation. 

Measurements of water deuteration through the different stages of star formation can therefore be used to trace the chemical evolution of water from the molecular cloud down to the protoplanetary disk. Observations of water are challenging due to the high abundance of water in the Earths atmosphere which makes the atmosphere opaque to prominent water emission lines. One solution is to observe water from space, as was done with the \textit{Herschel Space Observatory} \citep[see, e.g.,][]{dishoeck2011}. Observations from \textit{Herschel} have greatly expanded our knowledge of the different origins of water emission toward protostars, i.e., cold and warm envelope gas, shocks and outflows \citep[e.g.,][]{kristensen2012,coutens2013,visser2013}. 
Another approach is to target rarer water isotopologs such as H$_{2}^{18}$O or the deuterated isotopologs HDO and D$_2$O. These molecules have transitions which fall outside of the opaque water bands in the atmosphere and can be observed with ground based telescopes.
First attempts to constrain the HDO/H$_2$O ratio toward low-mass protostars utilized single-dish telescopes, thus observing a mixture of small- and large-scale emission and suffering from beam dilution. This led to a large discrepancy between measurements obtained with different telescopes and highlighted the importance of high spatial resolution to constrain the origin of the emission \citep[e.g.,][]{stark2004, parise2005, coutens2012}. 

\cite{jorgensen2010a} reported the first interferometric determination of the HDO/H$_2$O abundance ratio in the hot corino of NGC1333 IRAS 4B.
Since then water deuteration has been studied toward a number of low-mass Class 0 protostars using interferometers to resolve the water emission in the hot corino where $T > 100$~K and ice is entirely sublimated off the dust grains \citep{taquet2013observation, coutens2014, persson2014}.
These measurements reveals varying degrees of deuterium fractionation on different spatial scales toward Class 0 protostars. On larger spatial scales, in cold gas, a high degree of deuterium fractionation has been detected with HDO/H$_2$O and D$_2$O/HDO ratios of the order $\gtrapprox 10^{-2}$ in the gas-phase \citep{coutens2014}. Meanwhile, the water emission from the hot corino, i.e., small spatial scales, show lower HDO/H$_2$O ratios in the range $\sim10^{-4}$--$10^{-3}$. 
In comparison, the measured HDO/H$_2$O ratios in the Solar System range from as low as the local ISM value\footnote{[HDO/H$_2$O] = 2$\times$[D/H] and [D/H]$_\mathrm{ISM}$ = 2$\times10^{-5}$ \citep{prodanovic2010}.} ratio of $\sim 4\times10^{-5}$ to cometary values as high as [HDO/H$_2$O] $\sim 10^{-3}$; the Earth's D/H ratio, as measured from Vienna Standard Mean Ocean Water, is  D/H = $1.557\times10^{-4}$ \citep{deLaeter2003} which corresponds to HDO/H$_2$O $\sim 3\times10^{-4}$.

A possible explanation for the observed variation of the HDO/H$_2$O ratio on different spatial scales is that water and its deuterated version are not well mixed in the ice. The evolution of water deuteration during star formation and the effects of the layered ice structure have been the subject of recent modelling efforts \citep[e.g.,][]{cazaux2011, taquet2013model, furuya2016}. Combining chemical models and the available observations of the HDO/H$_2$O and D$_2$O/HDO ratios, \cite{furuya2016} proposed that water is primarily formed in the molecular cloud stage, before the dense core phase. This leads to a lower deuterium fractionation of  water which initially freezes out onto the interstellar dust grains and constitutes the bulk of the water ice reservoir. Later, in the dense prestellar core phase, the deuterium fractionation is enhanced as the temperature drops and the visual extinction increases. In this phase highly deuterated ice is formed on top of the existing water ice on grain mantles.
This scenario can explain the observed variation in the HDO/H$_2$O ratio on different spatial scales. On larger scales, the observed gas-phase HDO/H$_2$O ratio reflects the high deuterium fractionation of the outer layers of ice formed in the dense core phase as well as gas-phase synthesis of deuterated water through ion-neutral reactions in the cold outer regions of the envelope \citep{taquet2014, furuya2016}. In this region, the dust temperature is not sufficiently high to entirely sublimate the ice, and the inferred gas-phase abundance comes from a combination of the photodesorption of the outermost ice layers and gas-phase reactions with H$_2$D$^{+}$ which can form HDO and D$_2$O under these conditions. 
Meanwhile, on smaller scales the entire ice is sublimated off the grains and the ratio is lowered since the bulk of the water is formed with lower deuteration in the molecular cloud phase. Hence the ice abundances record the physical and chemical conditions of the different stages of star formation and the abundance of deuterated water could reflect the duration of the dense core phase. This in turn may depend on the local cloud conditions, since timescales of star formation is likely influenced heavily by external factors influencing the stability of the cloud, such as turbulence and radiation \citep[e.g.,][]{ward-thompson2007}.

Here we present observations of HDO and H$_{2}^{18}$O toward three isolated Class 0 protostars, L483, BHR71--IRS1, and B335. We classify these sources as isolated as they are not associated with any known cloud complexes. This is in contrast to the previously targeted protostars which are identified as part of star-forming regions like NGC1333 and Ophiucus and likely influenced by the dynamics within these stellar nurseries. The sources were selected to address if and how the deuterium fractionation of water varies as a function of the local cloud environment. 

BHR71--IRS1 is a Bok globule located in the Southern Coalsack dark nebulae with a bolometric luminosity $L_\mathrm{bol} \approx 15~L_\odot$ \citep{tobin2019}. It is part of a wide binary system, with the companion located at 3200 au at a distance $d\approx200$~pc. 
Lynds 483, commonly referred to as L483, is an isolated dense core which harbors the infrared Class 0 source IRAS 18148--0440. Traditionally, L483 was associated with the Aquila Rift region at an inferred distance of $200$~pc, however recent astrometry has revised the distance to the Aquila Rift up to $d\approx436\pm$9~pc \citep{ortiz-leon2018}. Subsequent analysis based on stellar extinction and parallaxes from Gaia DR2 has shown that L483 is in fact located at a distance of 200-250~pc, i.e., not a part of the Aquila Rift complex \citep{jacobsen2018}. We assume a distance of $200$~pc in this paper. At this distance the estimated bolometric luminosity is 10-13~$L_\odot$ \citep{tafalla2000,shirley2000}.
B335 is a Bok globule located at $d\approx100$~pc \citep{olofsson2009} and is the least luminous of our targets with a bolometric luminosity of $L_\mathrm{bol} \approx 0.72~L_\odot$ \citep{evans2015}. The central source is also known as IRAS 19347+0727 and identified as a low-mass Class 0 protostar. The central gas shows signs of infall and a rotational structure \citep[e.g.,][]{imai2019}.

The paper is structured as follows. The observations and calibration procedures are presented in Section \ref{sec:2}. Results are presented in Section \ref{sec:3}. The HDO/H$_2$O ratios are deduced and compared with previous measurements of water deuteration toward protostars in Section \ref{sec:4} along with a discussion of the implications. Finally, the results are summarized in Section \ref{sec:5} along with an outlook for the study of water deuteration and the impact of the local cloud environment.


   \begin{figure}
   \centering
   \includegraphics[width=\hsize]{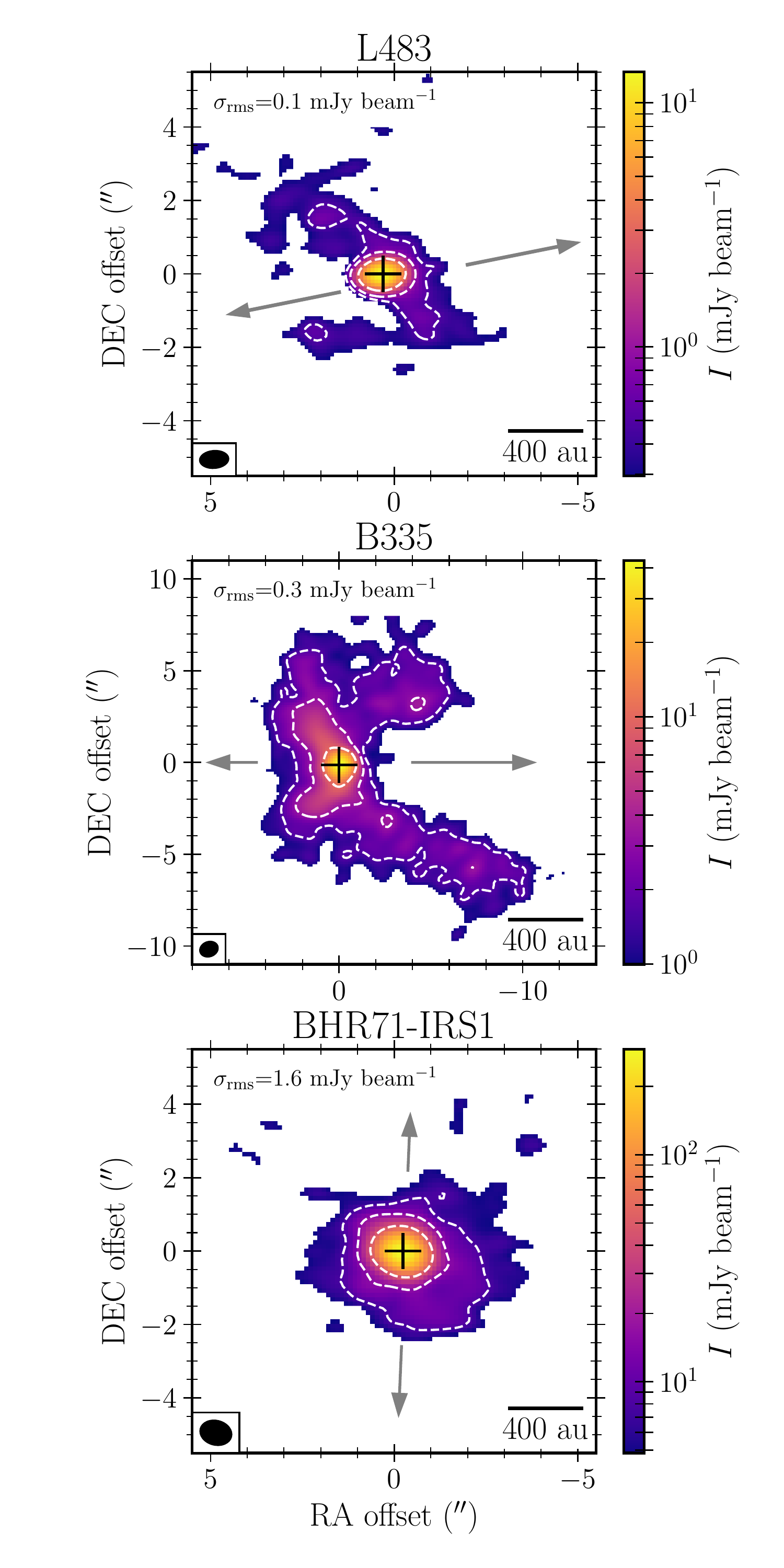}
      \caption{Continuum emission at 202.7~GHz toward the three sources. The map shows emission above 3$\sigma_\mathrm{rms}$ with white contours at 5$\sigma_{\mathrm{rms}}$, 10$\sigma_{\mathrm{rms}}$, and 30$\sigma_{\mathrm{rms}}$. The maps are presented on same linear scale. The black cross marks the peak location and the grey arrows indicate the potential direction of outflows which may be perturbing the dust distribution. For BHR71--IRS1 the grey arrows indicates the outflow direction presented by \cite{benedettini2017} and \cite{yang2017_BHR71}.
              }
         \label{fig:cont}
   \end{figure}

\section{Observations} \label{sec:2}
The low-mass embedded protostars L483, BHR71--IRS1, and B335 were observed with ALMA during Cycle 5 (PI Jes K. J{\o}rgensen, projectid: 2017.1.00693.S).
For L483 the observations were centered on $\alpha_{\mathrm{J}2000}$=18:17:29.9, $\delta_{\mathrm{J}2000}$=--04:39:39.6, for BHR71--IRS1 $\alpha_{\mathrm{J}2000}$=12:01:36.5, $\delta_{\mathrm{J}2000}$=--65:08:49.3, and for B335 $\alpha_{\mathrm{J}2000}$=19:37:00.9, $\delta_{\mathrm{J}2000}$=07:34:09.6. Information on the observation dates and calibration sources can be found in Table \ref{table:observations}.

One spectral setup targeted the HDO $3_{1,2}$--$2_{2,1}$ and $2_{1,1}$--$2_{1,2}$ transitions at 225.8967~GHz (LSB) and 241.5616~GHz (USB) respectively in the ALMA band 6.
Another spectral setup targeted the H$_{2}^{18}$O $3_{1,3}$--$2_{2,0}$ transition at
203.4075~GHz in Band 5.
Each spectral window contains 1920 channels with a width of 122~\mbox{kHz} (0.11 km s$^{-1}$). The source velocities, estimated from the HDO $3_{1,2}$--$2_{2,1}$ transition are $4.5$~km~s$^{-1}$, $-5.0$~km~s$^{-1}$, $7.9$~km~s$^{-1}$ for L483, BHR71--IRS1, and B335 respectively.

\begin{figure*}
   \resizebox{\hsize}{!}
            {\includegraphics{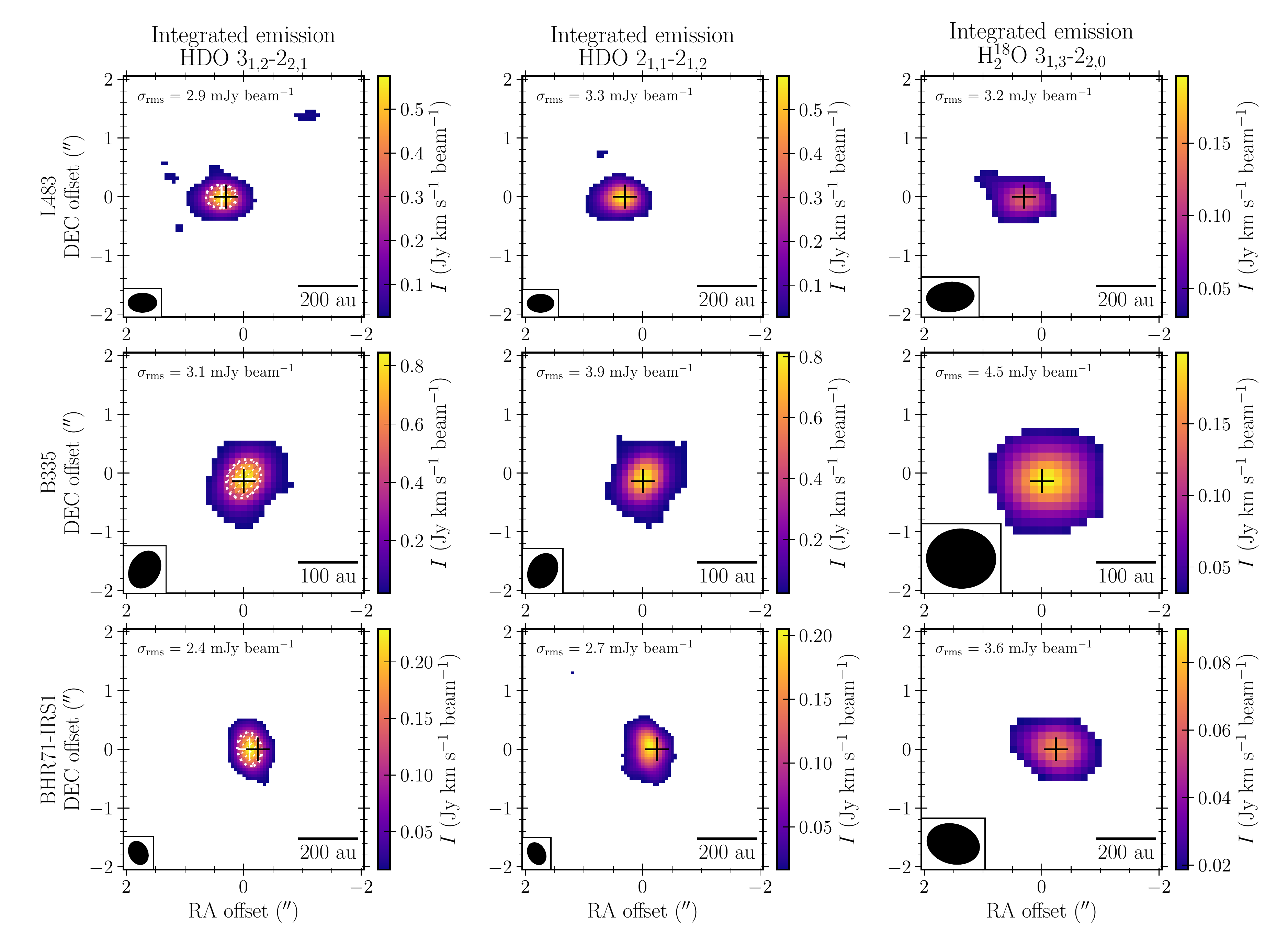}}
      \caption{Integrated emission for the targeted water transitions toward each of the sources. \emph{Left column:} HDO $3_{1,2}$--$2_{2,1}$ transition at 225.9~GHz. The white shaded regions show the FWHM extent of a 2D gaussian fitted to the data in the image plane and the white cross indicates the peak position of the fit.  \emph{Middle column:} HDO $2_{1,1}$--$2_{1,2}$ transition at 241.6~GHz. \emph{Right column:} H$_{2}^{18}$O $3_{1,3}$--$2_{2,0}$ transition at 203.4~GHz. \newline
      Emission below $5\sigma$ is not included, where $\sigma = \sigma_{\mathrm{rms}} \times N_\mathrm{channels}^{0.5} \times d\varv$; $d\varv$ is the channel width and $N$ is the number of collapsed channels. The black cross marks the 202.7~GHz continuum peak position toward which the spectra are extracted.}
         \label{fig:maps}
   \end{figure*}

Each dataset was pipeline-calibrated using {\sc casa 5.1} \citep{casa}. 
Phase self-calibration was performed using continuum channels for each dataset with {\sc casa 5.4}. 
For B335 a substantial improvement in the S/N ratio was achieved through self-calibration, while only marginal gains were achieved for BHR71-IRS1 and no gains for L483. For the latter we opted to use the pipeline product, since the self-calibrated data offered no improvements. For the self-calibrated sources we performed continuum subtraction using {\sc casa uvcontsub} before inversion. The images were deconvolved using the {\sc tclean} algorithm with a robust parameter of $-0.5$. 
For each source a continuum image was created at 202.7~GHz. The synthesized beam size range from $0\farcs4\times0\farcs3$ to $0\farcs7\times0\farcs5$ for the HDO spectral windows and $0\farcs8\times0\farcs5$ to $1\farcs2\times1\farcs0$ for the H$_{2}^{18}$O spectral window (see Fig. \ref{fig:maps}).



\begin{table*}
\centering\caption{Observation log.}             
\label{table:observations}      
\centering          
\begin{tabular}{l l c c c c c c}  
\hline \hline       
            \noalign{\smallskip}
Source &   Date & Phase Calibrator & Bandpass Calibrator & Max. baseline~(m) & N$_{\mathrm{antenna}}$ & ALMA Band\\  
            \noalign{\smallskip}
\hline                   
            \noalign{\smallskip} 
            L483  & 2017 March 11 &  J1743$-$0350 & J1751+0939  & 1100 & 42 & 6 \\
            L483  & 2017 August 27 & J1743$-$0350 & J1751+0939 & 768 & 44 & 5 \\
            BHR71--IRS1 & 2017 January 15 & J1147$-$6753 & J0904$-$5735 & 1797 & 46 & 6 \\
            BHR71--IRS1 & 2017 September 4 & J1147$-$6753 & J1107$-$4449 & 677 & 43 & 5 \\
            B335 & 2017 March 20 & J1955+1358 & J2025+3343 & 740 & 44 & 6 \\
           	B335 & 2017 August 27 & J1938+0448 & J2000$-$1748 & 759 & 45 & 5\\ 
            \noalign{\smallskip} 
\hline                                    
\end{tabular}
\end{table*}


\section{Results} \label{sec:3}
Figure \ref{fig:cont} shows the continuum emission at 202.7~GHz above 3$\sigma_{\mathrm{rms}}$ toward the three sources. All sources are clearly detected and the continuum structure is resolved on linear scales of $\sim 100$ au. Toward L483 the continuum emission is extended along the north-west to south-east diagonal potentially tracing the cavity walls of an outflow directed perpendicular to this direction. Toward B335 the dust emission extends far out from the central source with a notable lack of emission in the east-west direction, likely driven by outflows along this direction. Since B335 is located closer than the other two sources more of the envelope emission is filtered out by the interferometer. Comparing the continuum toward B335 with \cite{imai2016}, who observed the source at similar angular resolution and continuum wavelength, we see good agreement: the 10$\sigma$ contours (3 mJy~beam$^{-1}$) appear almost identical.
Toward BHR71--IRS1 the continuum appears more circular with no evidence of the known outflows perturbing the dust in the plane of the sky.

The HDO emission lines are identified using data from the Jet Propulsion Laboratory \citep[JPL,][]{JPL}, referencing \cite{HDO_ref}, while the H$_{2}^{18}$O transitional data originate from the Cologne Database for Molecular Spectroscopy \citep[CDMS,][]{CDMS}, with spectral details from \cite{h218o_ref}. All querying was done through the Splatalogue interface\footnote{\url{http://www.cv.nrao.edu/php/splat/}}.

\begin{figure*}
   \resizebox{\hsize}{!}
            {\includegraphics{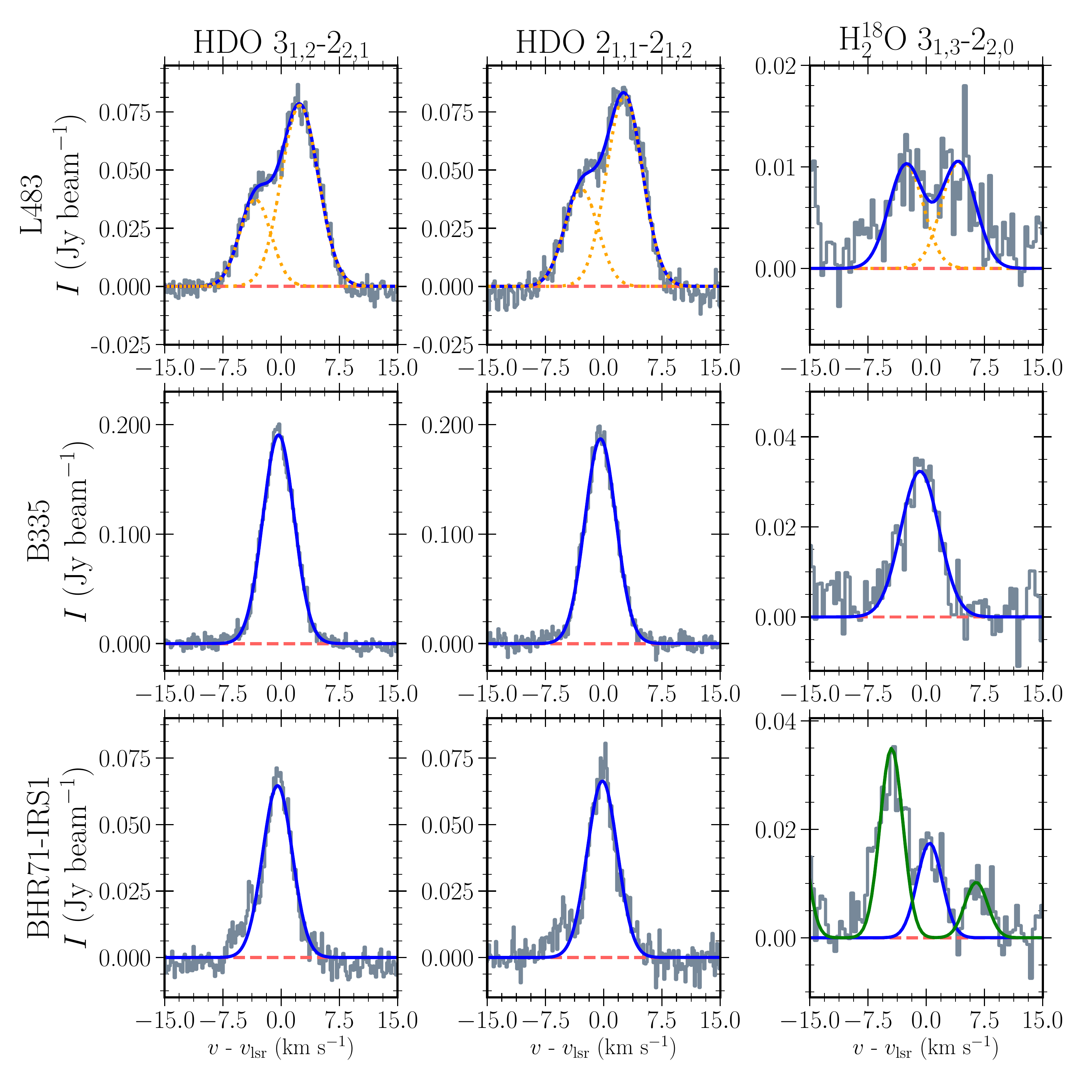}}
      \caption{Continuum-subtracted spectra for the three target transitions, extracted toward the continuum peak. Blue lines represent the gaussian fits. For BHR71--IRS1, dimethyl ether lines (green) are fitted and subtracted from the spectrum before the H$_{2}^{18}$O line (blue) is fitted. The H$_{2}^{18}$O lines have been rebinned by a factor of two for clarity.}
         \label{fig:fitted_lines}
   \end{figure*}

The targeted water isotopologs are detected toward each of the sources: the HDO transitions are detected with high signal to noise (SNR $> 10$) while the H$_{2}^{18}$O lines are slightly weaker detections (SNR $\sim 5-10$, Fig. \ref{fig:maps}).
The targeted emission lines are presented in Figure \ref{fig:fitted_lines} along with fitted gaussian profiles. The H$_{2}^{18}$O data are rebinned by a factor of two for clarity, however this does not influence the abundances and HDO/H$_2$O ratios derived in this paper. 

Each spectrum is extracted toward the 202.7~GHz continuum peak of the source (see Fig. \ref{fig:cont}). For the spectral windows centered on the HDO transitions the image cube is convolved with a gaussian kernel using the {\sc imsmooth} function in {\sc casa}. The size of the gaussian kernel is determined such that the shape of the beam in the resulting image matches the larger beam of the H$_{2}^{18}$O image cube at 203~GHz, allowing for direct comparison between the extracted spectra (see Fig. \ref{fig:maps} for beam sizes before the convolution).
Toward B335 and BHR71--IRS1 the emission lines are fitted using a single gaussian profile. Toward L483 the emission lines show two components: a brighter redshifted component and weaker blueshifted component. In this case two gaussian profiles were used to fit the lines. 
The double-peaked line profiles are consistent with previous high-resolution observations of hot corino species toward L483 presented in \cite{oya2017} and \cite{jacobsen2018} who independently found evidence of infalling rotating motion in the hot corino region toward the source.

Toward BHR71--IRS1 there is a slight hint of a broader outflow component in the wings of the emission lines, but this component is too weak to influence the fitted profile significantly. No extended outflow structure is seen when imaging the line-wings of any of the water transitions.
The H$_{2}^{18}$O line is partialy blended with dimethyl ether (CH$_3$OCH$_3$) toward BHR71-IRS1. In this case, six transitions of CH$_3$OCH$_3$ are fitted with gaussian profiles using a fixed FWHM and system velocity and the fit is subsequently subtracted from the spectrum to remove the blending before the H$_{2}^{18}$O line is fitted.

In Appendix \ref{app:velocity} we present velocity field for the three transitions toward each source. These maps confirm that no high-velocity outflow components contribute to the observed emission for the emission lines in question. Toward L483 the velocity field is consistent with the results of \citet{jacobsen2018}.

Figure \ref{fig:maps} shows integrated emission maps for H$_{2}^{18}$O and HDO toward the three sources. The detected emission originates from the central $\sim 100$ au for each source, consistent with the emission originating in the hot corino ($T>100$~K). In both cases the central emission is unresolved and the slight variation in the extent of the emitting region can be explained by the differences in synthesised beam size. The peak positions of the continuum emission and water emission overlap down to scales of the synthesised beam size. 

Since we aim to determine the gas-phase abundances of HDO and H$_2^{18}$O in the hot corino, it is essential that the observed emission lines originate from this region. We are confident this is the case for several reasons.
The emission is compact, originating from scales of less than 100~au. This is confirmed by fitting two-dimensional gaussian profiles to the emission maps. Here the FWHM extent on linear scales ranges from 50~au to 100~au. The fitted 2D profiles for the HDO $3_{1,2}-2_{2,1}$ transition toward the three sources are shown in the left column of Figure \ref{fig:maps}.
Additionally, the upper energy levels of the detected transitions lie around 100--200~K and the transitions are thus not easily excited in the cold outer envelope.
Lastly, the line widths (FWHM $\lesssim 6~\mathrm{km}~\mathrm{s}^{-1}$) of the emission lines are consistent with emission from the hot corino, with little or no evidence of outflow emission blending as described above and the emission is not extended along the known outflow directions, as in NGC1333 IRAS 2A \citep{persson2012}.

\begin{table*}
\caption{Fit parameters for the targeted HDO and H$_{2}^{18}$O transitions.}             
\label{table:lines}      
\centering 
\smallskip \smallskip

\begin{tabular}{c c c c c c}        
\hline\hline                 
            \noalign{\smallskip} 

Species & $\nu_\mathrm{rest}$ (\mbox{GHz}) & $F_{\nu}^{\mathrm{peak}}$ \mbox{(mJy beam$^{-1}$)} & FWHM (km~s$^{-1}$) & $\varv_{\mathrm{lsr}}$ (km~s$^{-1}$) & $N$ (cm$^{-2}$) \\    
\hline                        
            \noalign{\smallskip} 
  \multicolumn{6}{c}{\it L483} \\
\hline

   HDO & 225.896720$^{a}$ & $37\pm4$ & $4.8\pm0.3$ & $1.7\pm0.1$ & ($1.7\pm0.2$)$\times10^{15}$ \\  
   HDO & 225.896720$^{b}$ & $78\pm8$ & $5.6\pm0.2$ & $7.4\pm0.1$ & ($4.3\pm0.4$)$\times10^{15}$ \\ 
   HDO & 241.561550$^{a}$ & $42\pm5$    & $5.1\pm0.2$ & $2.1\pm0.2$ & ($1.8\pm0.2$)$\times10^{15}$ \\
   HDO & 241.561550$^{b}$ & $82\pm8$ & $5.5\pm0.4$ & $7.7\pm0.2$ & ($3.8\pm0.4$)$\times10^{15}$ \\

   H$_{2}^{18}$O & 203.407520$^{a}$ & $10.2\pm1.5$   & $5.2\pm1.0$ & $2.4\pm0.4$ & ($2.3\pm0.6$)$\times10^{15}$ \\
   H$_{2}^{18}$O & 203.407520$^{b}$ & $10.5\pm1.5$   & $5.2\pm1.0$ & $9.2\pm0.4$ & ($2.4\pm0.6$)$\times10^{15}$\\
            \noalign{\smallskip} 
\hline                                   
            \noalign{\smallskip} 
  \multicolumn{6}{c}{\it B335} \\
\hline

   HDO & 225.896720 & $190\pm19$ & $4.74\pm0.04$ & $7.97\pm0.16$ & ($3.1\pm0.3$) $\times 10^{15}$ \\ 
   HDO & 241.561550 & $187\pm19$    & $4.71\pm0.05$ &  $7.89\pm0.15$ & ($2.6\pm0.3$) $\times 10^{15}$ \\
   H$_{2}^{18}$O & 203.407520 & $32\pm4$   &  $5.78\pm0.36$ & $7.49\pm0.14$ & ($2.9\pm0.4$) $\times 10^{15}$ \\
            \noalign{\smallskip} 
\hline                                   
            \noalign{\smallskip} 
  \multicolumn{6}{c}{\it BHR71--IRS1} \\
\hline

   HDO & 225.896720 & $65\pm7$ & $4.51\pm0.16$ &  -$4.97\pm0.05$ & ($1.9\pm0.2$) $\times 10^{15}$ \\  
   HDO & 241.561550 & $66\pm7$ & $4.57\pm0.15$ &  -$4.67\pm0.05$ & ($2.2\pm0.2$) $\times 10^{15}$ \\
   H$_{2}^{18}$O & 203.407520 & $17\pm3$ & $3.80\pm0.48$ & -$4.07\pm0.20$ & ($2.0\pm0.4$) $\times 10^{15}$\\
            \noalign{\smallskip} 
\hline                                   
\end{tabular}
\tablefoot{$F_\mathrm{\nu}$ includes 10\% calibration uncertainty. FWHM uncertainty is determined as the maximum between the uncertainty in the fitted gaussian and the channel width. Column densities $N$ were calculated assuming optically thin emission from a gas in LTE at 124~K. Furthermore, the column densities assume that the emission fills the beam. Toward L483, ${}^a$ denotes the weaker blueshifted component and ${}^b$ the brighter redshifted component.\\}
\end{table*}

   \begin{figure*}
   \resizebox{\hsize}{!}
            {\includegraphics{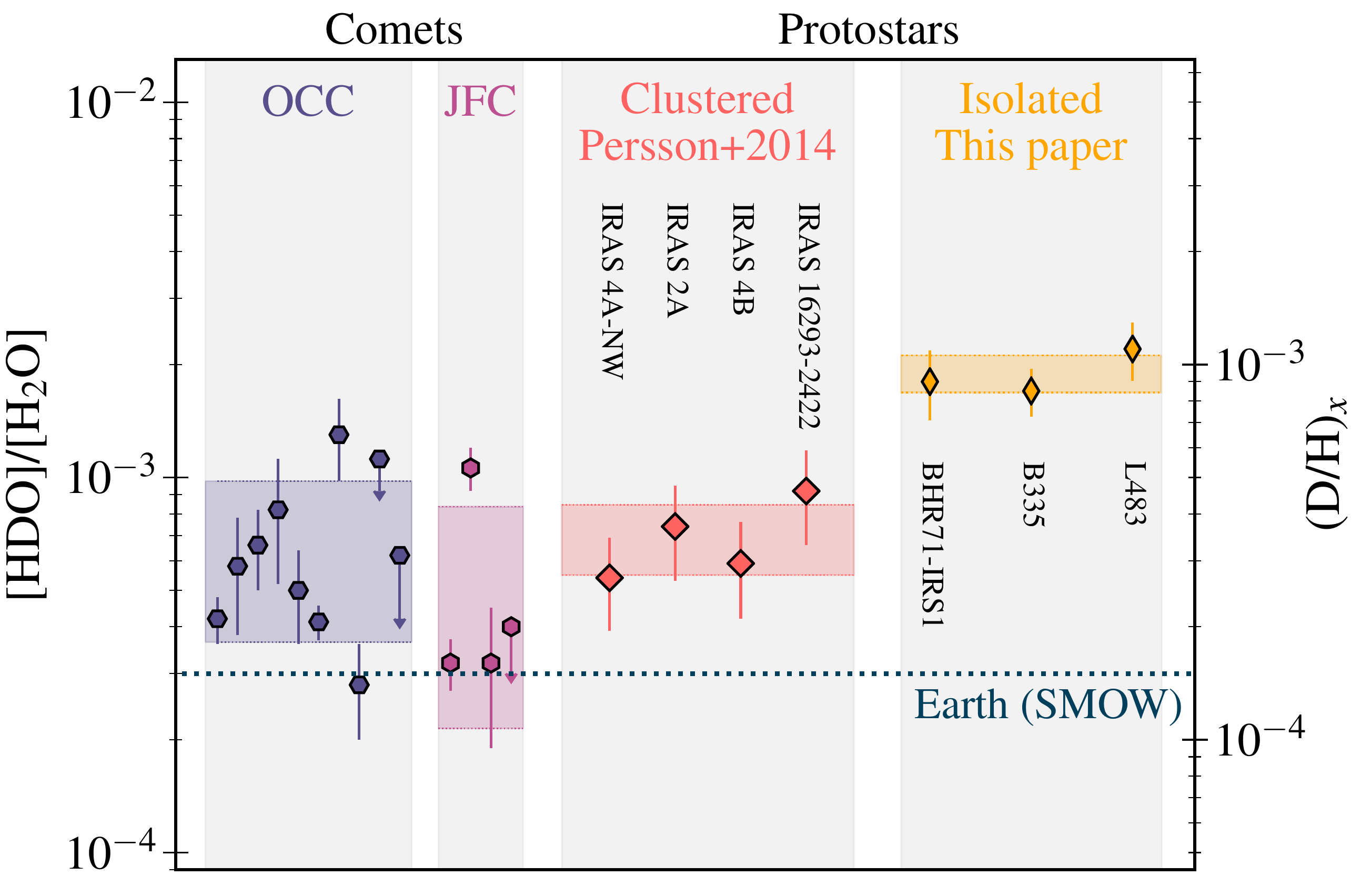}}
      \caption{Comparison between the D/H ratio and HDO/H$_2$O ratio for comets in the Solar System and hot corino observations toward Class 0 protostars. Values for IRAS 16293--2422, IRAS 2A, and IRAS 4B are from \cite{persson2014}, while IRAS 4A--NW has been adjusted from the value quoted in the paper as a mistake in the data analysis was discovered, which enhanced the HDO abundance by a factor of $\sim$2. Errorbars show 1$\sigma$ uncertainties. For the isolated sources the uncertainty is based on statistical errors from the fitted gaussian profiles and a flux calibration uncertainty of 10\%. On the right axis the corresponding D/H ratio is shown.
      The references for the Oort Cloud Comets (OCC) and Jupiter Family Comets (JFC) can be found in Appendix \ref{app:table}. The colored regions show the standard deviation for each class of objects. Note that the HDO/H$_2$O ratio for the protostars are derived from observations of HDO and H$_{2}^{18}$O while some cometary values are derived from other proxies for the D/H ratio.}
         \label{fig:objects_comparison}
   \end{figure*}

\section{Analysis and discussion}  \label{sec:4}
\subsection{Estimating the column densities of HDO and H$_{2}^{18}$O and deriving the HDO/H$_2$O ratio}
We estimated the column densities of HDO and H$_{2}^{18}$O considering optically thin emission and  local thermodynamic equilibrium (LTE) with an excitation temperature of 124~K. The choice of analysis was motivated by the aim to compare the present results with previous measurements presented in \cite{persson2014} and \cite{coutens2014} who determined the water deuteration in the hot corino toward a number of sources using the same method and excitation temperature. Choosing the same methodology, we can directly compare the water deuteration under similar assumptions. 
The results are summarized in Table \ref{table:lines}. The estimated column densities range from $(2-7)\times10^{15}$ cm$^{-2}$, comparable to the estimates of \cite{persson2014} where the HDO column densities range from $1.5\times10^{15}$ cm$^{-2}$ for the faintest source, IRAS 4B, to $1.2\times10^{16}$ cm$^{-2}$ for IRAS 2A. 

The estimated HDO/H$_2$O ratios for L483, B335, and BHR71--IRS1 are ($2.2\pm0.4$)$\times 10^{-3}$, ($1.7\pm0.3$)$\times 10^{-3}$, and ($1.8\pm0.4$)$\times 10^{-3}$, respectively. Uncertainties are derived from the statistical uncertainties of the fitted gaussian profiles with an additional 10\% uncertainty on the flux calibration. From the column density of H$_{2}^{18}$O we infer the H$_2$O water column densities by assuming the Galactic oxygen isotope ratio of $^{16}$O/$^{18}$O$ = 560$ \citep{wilson1994}. The HDO column density is determined as the weighted average of the two transitions. 
The HDO/H$_2$O ratios toward the three sources show little scatter and are well within the uncertainties of one another which suggests similar chemical evolution of water in these systems. 

The assumption of a fixed excitation temperature for the three sources only has a moderate effect on the estimated HDO/H$_2$O ratios. Calculations presented by \cite{persson2014} and \cite{jorgensen2010a} have shown that varying the excitation temperature in the range 50--300~K has limited influence on the HDO column densities in the LTE approximation. We confirm this in Appendix \ref{app:tex}, where the HDO/H$_2$O ratio is calculated for the measured line strengths and excitation temperatures in the range 30~K to 300~K. Toward L483, an excitation temperature $T_\mathrm{ex}\leqslant$ 60~K is needed to bring the HDO/H$_2$O ratio within the range of the four clustered sources reported in \cite{persson2014} while B335 and BHR71--IRS1 require $T_\mathrm{ex}\leqslant$ 70~K. Such low excitation temperatures in the inner $\sim50$-$100$~au are not expected. 
Since we observe two transitions of HDO for all three sources we can calculate the excitation temperature for each source under the assumption of LTE. The  excitation temperatures are 127$\pm24$~K, 174$\pm59$~K, and 103$\pm$21~K for L483, B335, and BHR71--IRS1; hence the choice of 124~K as the excitation temperature is consistent with the HDO line strengths toward the three sources. Meanwhile, the derived excitation temperature for HDO is inconsistent with the lower excitation temperatures needed to bring the HDO/H$_2$O ratio down the values similar to the clustered sources, i.e., 60~K for L483 and 70~K for B335 and BHR71--IRS1. 
Computing the HDO/H$_2$O ratio for the estimated excitation temperatures yields ($2.2\pm0.4$)$\times 10^{-3}$ for L483, ($2.1\pm0.3$)$\times 10^{-3}$ for B335, and ($1.6\pm0.3$)$\times 10^{-3}$ for BHR71--IRS, all well within the error bars. Using the estimated excitation temperatures rather than the fixed value does not affect the conclusions presented here.

The difference between optically thin LTE calculations, similar to those presented here, and more advanced modelling for comparable observations of HDO and H$_2^{18}$O have been studied in previous works to determine how well the former assumptions work in this regime. \cite{persson2014} ran radiative transfer models for IRAS 16293--2422 and IRAS 2A and found that the water column densities were consistent with those derived from optically thin LTE calculations. Similarly, \cite{coutens2014} ran non-LTE {\sc radex} calculations and found column densities consistent with the LTE calculations. These results suggest that the water emission originating in the hot corino of these low-mass Class 0 protostars is well approximated by LTE calculations and that the emission is not optically thick. We note that if the densities in the inner envelopes of the isolated protostars are lower than the clustered counterparts then the emission from the water isotopologs could be sub-thermal and the LTE approximation invalid. We consider it unlikely to be the case; observational estimates of the overall envelope masses and bolometric luminosities for L483 and B335 are comparable to those of IRAS 2A, IRAS4A, and IRAS4B \citep[see, e.g.,][]{shirley2002, jorgensen2007, kristensen2012}. With comparable envelope masses and luminosities it is unlikely that the inner density profile, i.e. on hot corino scales, would differ substantially between the isolated and clustered sources. 

The advantage of more advanced models, as opposed to the LTE approach adopted here, is often diminished by the uncertainties of the physical source parameters such as the kinematics, the density profile, protostellar parameters etc. 
Attempting more advanced modelling may thus offer little improvement over the LTE approach and makes direct comparison between sources more challenging.

\subsection{Water deuteration: Comparison with existing observations}
Figure \ref{fig:objects_comparison} shows the calculated HDO/H$_2$O ratios for the three isolated protostars along with existing data for a number of low-mass embedded protostars in clustered star-forming regions and cometary values from the Solar System. The protostellar values in Figure \ref{fig:objects_comparison} are all derived from interferometric observations with high spatial resolution to probe the hot corino emission where the ice is sublimated entirely off the dust grains.
The three sources observed in this work have a similar degree of deuterium fractionation with HDO/H$_2$O ratios in the range $(1.7-2.2)\times10^{-3}$. Meanwhile, the four previously observed protostars lie in the range $(5.5-9.2)\times10^{-4}$. This suggests that the sources presented here, which are all isolated protostars, have a distinct chemical history compared to the clustered protostars previously targeted.
Comets for which the D/H ratio have been determined generally show lower values than the isolated protostars as seen in Figure \ref{fig:objects_comparison}. Meanwhile, the clustered protostars show reasonable agreement with the Oort Cloud Comets, which have led to the suggestion that comets form from gas that is chemically similar to the gas observed in hot corinos, with little or no processing after this stage \citep{persson2014}. Assuming this is the case, and that the D/H ratio of comets have not changed significantly after their formation, this would suggest that the Solar System was formed in a clustered region of star formation. Such a scenario is also supported by evidence such as the abundance of short-lived radionuclides, high-eccentricity orbits of small Solar System bodies, and low occurrence rate of isolated protostars \citep{adams2010}.


The apparent differentiation between clustered and isolated protostars can be understood in the framework for water formation and deuterium fractionation proposed by \cite{furuya2016}. They propose that water is primarily formed in the molecular cloud with limited deuteration. Later on, in the dense-core phase, deuteration is enhanced due to the low temperature and high shielding leading to the freeze-out of CO and a higher D/H ratio in the gas-phase. In this scenario, the duration of the dense-core phase determines the amount of deuteration, with a longer dense-core phase resulting in an enhanced deuterium enrichment in the ice. Conversely, a prolonged molecular cloud phase could decrease the deuterium fractionation and the HDO/H$_2$O ratio is thus related to the ratio of the life-time of the two stages.

The star-formation process is recognized to be heterogenous in nature, with some stars born in dense clusters while others are formed in relatively isolated regions of molecular clouds \citep[e.g.,][]{ward-thompson2007}. Stars born in dense clusters are likely to collapse on shorter timescales; the mean free-fall time scales with the inverse square root of the mean density $\rho_0$, $t_{\mathrm{ff}} = (3\pi/32G\rho_{0})^{1/2}$ \citep[e.g.,][]{padoan2014, krumholz2014}. Furthermore, external pressure from nearby massive stars in the local cloud region can trigger and potentially accelerate the collapse process in dense star forming regions. Accordingly, isolated protostars could experience a longer dense-core phase, enhancing the deuterium fractionation and potentially also enhancing the abundance of complex organic molecules formed from CO ice. 
The isolated protostars presented here all exhibit a higher deuterium fractionation than comparable counterparts in more dense, star-forming regions, in agreement with the theoretical expectation as outlined above.
Assuming that the water formation processes are understood, this enhanced deuterium fractionation can imply two things: either the timescale of the dense-core phase is longer for the isolated cores or the molecular cloud phase, where most H$_2$O is formed, is shorter for clustered star formation. The latter option does not appear likely since both theoretical and observational data imply that isolated cores do not collapse more easily than their clustered counterparts \citep{ward-thompson2007}.
Given that the isolated protostars presented here exhibit higher deuterium fractionation compared to clustered protostars, this implies that the ratio between the duration of the dense-core phase and the molecular cloud phase is higher for these sources. 

Another possible explanation for the differences between the HDO/H$_2$O ratios toward clustered and isolated protostars could be variations in the temperatures between the different regions. For example, higher initial gas temperatures in the clustered regions, e.g., due to irradiation by neighboring young stars, would also reduce the efficiency of deuteration enrichment in these regions. Further, observational constraints and numerical studies are needed to address the relative importance of the collapse time-scales and gas temperatures.
 

The observations presented here support the hypothesis that the local cloud environment influences the early physical evolution and ultimately the chemistry of young stellar systems. A consequence of this is that the water deuteration can be an important proxy for both the chemical and the physical history of protostars.

Regarding the question of inheritance or local processing of water during star formation, the results presented here favor the inheritance scenario, at least at the earliest protostellar phase, since all three isolated protostars show similar HDO/H$_2$O ratios. The lack of pronounced variation between the sources indicate a similar physical and chemical evolution for the systems, with little impact from local variations in, e.g., protostellar luminosity or accretion bursts. This conclusion is compatible with previous studies which modeled the water evolution from the collapse of an isolated pre-stellar cores to circumstellar disk \citep{visser2009, cleeves2014, drozdovskaya2016, furuya2017}. 

\section{Summary and outlook}  \label{sec:5}
In this paper we present the first ALMA Band 5 observations of the H$_{2}^{18}$O $3_{1,3}$--$2_{2,0}$ transition toward three isolated low-mass Class 0 protostars. Combined with observations of the HDO $3_{1,2}$--$2_{2,1}$ and $2_{1,1}$--$2_{1,2}$ transitions we have determined the HDO/H$_2$O ratio for the sources and compared with previous determinations of the water deuteration toward low-mass Class 0 protostars.
   \begin{enumerate}
   	  \item The targeted water emission is detected in the hot corino toward the targeted sources with a high S/N on angular scales of 0\farcs3-1\farcs1 corresponding to linear scales of $\sim$50-150~au. The column densities of HDO and H$_2^{18}$O have been determined assuming optically thin emission under local thermodynamic equilibrium with an excitation temperature of 124 K. 
   	  \item  From the estimated column densities the derived HDO/H$_2$O ratios for L483, B335, and BHR71--IRS1 are ($2.2\pm0.4$)$\times 10^{-3}$, ($1.7\pm0.3$)$\times 10^{-3}$, and ($1.8\pm0.4$)$\times 10^{-3}$, respectively.
      \item  The three isolated protostars show a factor of $\sim2$ higher deuterium fractionation than the previously targeted sources, which are associated with larger cloud complexes. This observed differentiation in water deuteration is significant and can be explained by variations in the collapse timescales or initial gas temperatures depending on the cloud environment. In clustered regions external perturbations from nearby stars may either accelerate the collapse process through turbulence or heat the gas, leading to a lower deuterium fractionation. If this is the case then the degree of deuterium fractionation correlates with the local cloud environment, providing a new proxy for the early evolutionary history of young stars.
      \item The similarity of the HDO/H$_2$O ratio toward the three isolated protostars could indicate that little processing of the water has occurred from cold cloud to hot core and suggests that the conditions in the dense core phase, before the onset of the collapse, determines the deuterium fractionation at later stages.

\end{enumerate}

These observations present the first measurements of water deuteration targeting specific cloud environments and have nearly doubled the number of protostars for which the hot corino water deuteration has been measured. The observations indicate that isolated protostars have a distinct chemical history and further exploration of the relationship between cloud environment and deuterium fractionation could strengthen our understanding of the physical and chemical evolution during star formation.
A natural progression is to determine the water deuteration toward more clustered or high-mass protostars, which should show lower fractionation levels according to the chemical evolution outlined above.
Another option is to target doubly deuterated water, D$_2$O. At this stage the D$_2$O column density has only been determined toward the hot corino region of one source, NGC 1333 IRAS 2A \citep{coutens2014}. Expanding the number of sources for which the D$_2$O/HDO ratio is measured would test if the trend shown in Fig. \ref{fig:objects_comparison} is real. Should the trend not be present for the doubly deuterated water isotopolog then we are missing important details in the current chemical models.

A third option to constrain the importance of the local cloud environment for the chemical evolution is to determine the deuteration of other molecules, such as methanol, in the hot corino. If the relationship between local cloud environment and molecular deuteration proposed here is correct, then the deuteration should be enhanced for all molecules.

In parallel with observational efforts, numerical modeling is needed to improve our understanding of the effects of the local cloud environment on the chemistry of young stellar systems. So far, little work has been done to model the chemical evolution in the context of a dynamic molecular cloud environment, i.e., collapse models which include the influence of the surrounding cloud environment with the local differences in temperature, density, UV-field, and turbulence. Such modeling would strengthen our understanding of the link between chemistry, particularly the deuterium fractionation, and the local cloud environment.

\begin{acknowledgements}
We thank the anonymous referee for valuable comments, which improved the manuscript.
This paper makes use of the following ALMA data: ADS/JAO.ALMA\#2017.1.00693.S. ALMA is a partnership of ESO (representing its member states), NSF (USA) and NINS (Japan), together with NRC (Canada), NSC and ASIAA (Taiwan), and KASI (Republic of Korea), in cooperation with the Republic of Chile. The Joint ALMA Observatory is operated by ESO, AUI/NRAO and NAOJ.
	The group of JKJ acknowledges support from the European Research Council (ERC) under the European Union's Horizon 2020 research and innovation programme (grant agreement No 646908) through ERC Consolidator Grant "S4F". Research at the Centre for Star and Planet Formation is funded by the Danish National Research Foundation. Astrochemistry in Leiden is supported by the Netherlands Research School for Astronomy (NOVA). A.C. postdoctoral grant is funded by the ERC Starting Grant 3DICE (grant agreement 336474).
\end{acknowledgements}

%
%

\bibliographystyle{aa}
\bibliography{isolated_cores.bib}

\begin{appendix} 

\section{Table of D/H and HDO/H$_2$O ratios}\label{app:table}
Values reported in Figure \ref{fig:objects_comparison} are presented in Table \ref{table:deuteration} along with references and information of the tracers used to determine the water deuterium fractionation.
\begin{table*}
\caption{Measured HDO/H$_2$O and D/H ratios for comets and protostars.}             
\label{table:deuteration}      
\centering 
\smallskip \smallskip

\begin{tabular}{l c c c}        
\hline\hline                 
            \noalign{\smallskip} 

Object & HDO/H$_2$O ($\times10^{-4}$) &  Tracers & Reference \\
\hline                        
            \noalign{\smallskip} 
  \multicolumn{4}{c}{\it Oort Cloud Comets} \\
              \noalign{\smallskip} 
              
\hline
1/P Halley & $4.2\pm0.6$ & H$_2$DO$^+$, H$_3$O$^+$ & 1 \\
C/1996 B2 Hyatuake & $5.8\pm2.0$ & HDO, H$_2$O & 2 \\
C/1995 O1 Hale-Bopp & $6.6\pm1.6$ & HDO, H$_2$O & 3 \\
C/2007 B3 Lulin & $<11.2$ &  HDO, H$_2$O & 4 \\
8P/Tuttle & $8.2\pm3.0$ & HDO, H$_2$O & 5 \\
C/2009 P1 Garradd & $4.12\pm0.44$ &  HDO, H$_{2}^{18}$O, H$_2$O & 6 \\
C/2002 T7 LINEAR & $5.0\pm1.4$ &  OD, OH, $^{18}$OH & 7 \\
153P Ikeya-Zhang & $<5.6\pm0.6$ &  HDO, H$_{2}^{18}$O & 8 \\
C/2012 F6 Lemmon & $6.5\pm1.6$ & HDO, H$_{2}^{18}$O, H$_2$O & 9 \\
C/2014 Q2 Lovejoy & $1.4\pm0.4$ & HDO, H$_{2}^{18}$O, H$_2$O & 9 \\
\hline                        
            \noalign{\smallskip} 
  \multicolumn{4}{c}{\it Jupiter Family Comets} \\
\hline
45P Honda-Mrkos-Pajdusakov (HMP) & $<4.0$ &  HDO, H$_{2}^{18}$O, H$_2$O & 10 \\
103P Hartley 2 & $3.2\pm0.5$  & HDO, H$_{2}^{18}$O, H$_2$O & 11 \\
67/P Churyumov-Gerasimenko & $10.6\pm1.4$  & HDO, HD$^{18}$O, H$_2$O, H$_{2}^{18}$O & 12 \\
46P/Wirtanen & $3.2\pm1.3$ &  HDO, H$_{2}^{18}$O & 13 \\

\hline                        

            \noalign{\smallskip} 
  \multicolumn{4}{c}{\it Clustered protostars} \\
\hline
NGC1333 IRAS 4A-NW & $5.4\pm1.5$ & HDO, H$_{2}^{18}$O & 14$^{a}$ \\
NGC1333 IRAS 2A & $7.4\pm2.1$ & HDO, H$_{2}^{18}$O & 14 \\
NGC1333 IRAS 4B & $5.9\pm2.6$ & HDO, H$_{2}^{18}$O & 14 \\
IRAS 16293--2422 & $9.2\pm2.6$ &  HDO, H$_{2}^{18}$O & 14 \\
\hline                        
            \noalign{\smallskip} 
  \multicolumn{4}{c}{\it Isolated protostars} \\
\hline
BHR71--IRS1 & $18\pm4$ & HDO, H$_{2}^{18}$O & 15 \\
B335 & $17\pm3$ & HDO, H$_{2}^{18}$O & 15 \\
L483 & $22\pm4$ & HDO, H$_{2}^{18}$O & 15 \\
	\noalign{\smallskip} 
	\hline                        
\end{tabular}
\tablefoot{Protostars only include interferometric observations of hot corino emission toward low-mass Class 0 protostars. Conversions between HDO/H$_2$O and D/H assumes the statistical ratio HDO/H$_2$O = $2 \times$D/H. ${}^{a}$ Note that the value for IRAS 4A--NW has been adjusted from the value derived in \cite{persson2014} since a mistake in the data analysis was discovered, which enhanced the HDO abundance by a factor of $\sim$2.}
\tablebib{
(1) \citet{brown2012}; (2) \citet{bm1998}; (3) \citet{meier1998}; (4) \citet{gibb2012}; (5) \citet{villanueva2009}; (6) \citet{bm2012}; (7) \citet{hutsemekers2008}; (8) \citet{biver2006}; (9) \citet{biver2016}; (10) \citet{lis2013}; (11) \citet{hartogh2011}; (12) \citet{altwegg2015}; (13) \citet{lis2019}; (14) \citet{persson2014}; (15) This paper.
   }
\end{table*}

\section{Computing HDO/H$_2$O as a function of excitation temperature in the LTE approximation}\label{app:tex}
 
\begin{figure*}
   \resizebox{\hsize}{!}{\includegraphics{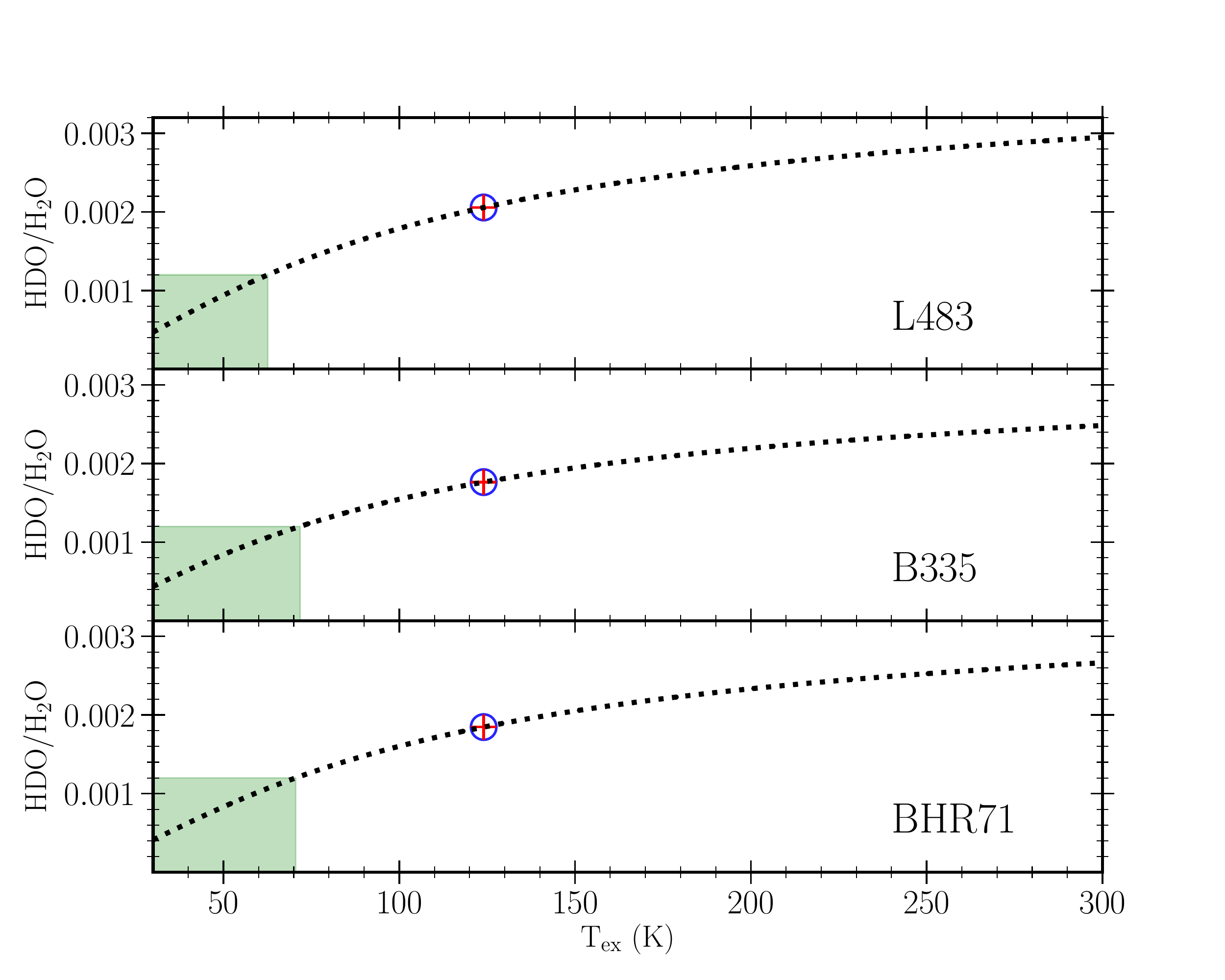}}
      \caption{HDO/H$_2$O ratio for the three sources assuming excitation temperatures in the range 30--300~K and optically thin LTE emission. The green shaded area show the upper limit for the HDO/H$_2$O ratio for the clustered protostars presented in \cite{persson2014}. The blue circle marks the value at the assumed excitation temperature of 124~K value.}
         \label{fig:tex}
   \end{figure*}

In Figure \ref{fig:tex} we show the HDO/H$_2$O ratio as a function of excitation temperature in the range 30--300~K under the assumption of optically thin LTE emission. All values are calculated using the line strengths from the fitted line profiles presented in Figure \ref{fig:fitted_lines}. Evidently the exact choice of excitation temperature in the regime 100--200~K has limited effect on the corresponding HDO/H$_2$O ratio.
The green shaded region indicates the upper limit for the clustered protostars in \cite{persson2014}, HDO/H$_2$O = $1.18\times10^{-3}$ for IRAS 16293--2422. For L483 an excitation temperature of $\sim 60$~K is needed to lower the HDO/H$_2$O to $1.18\times10^{-3}$ while the threshold for B335 and BHR71--IRS1 lies around 70~K.

\section{Velocity fields for HDO and H$_{2}^{18}$O}\label{app:velocity}
Moment 1 velocity fields computed in {\sc casa} to verify that no high-velocity components are present for any of the targeted water transitions.
   \begin{figure*}
   \resizebox{\hsize}{!}{\includegraphics{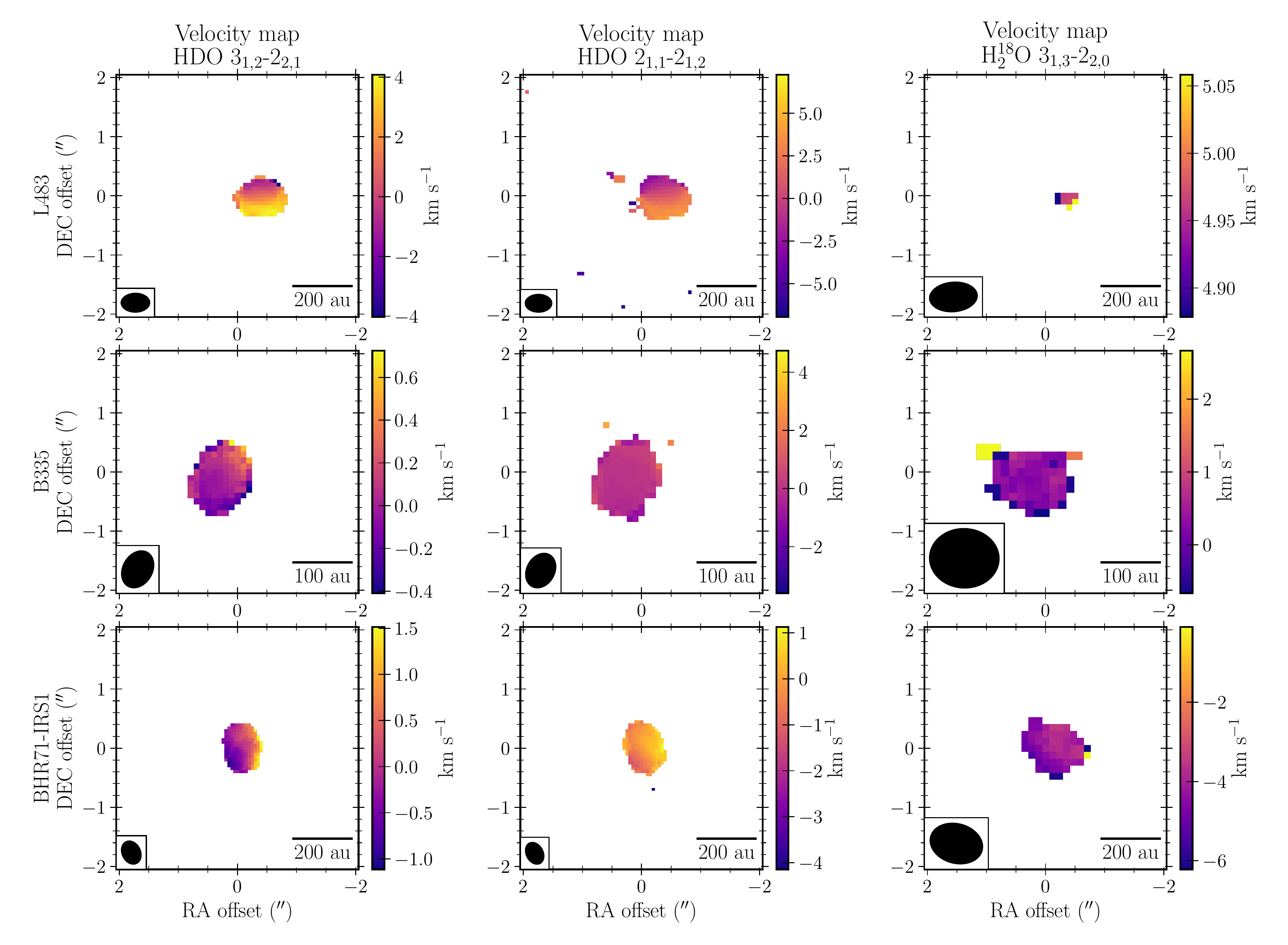}}
      \caption{Moment 1 velocity fields toward each of the sources for the three water transitions. The maps show no signs of outflow emission contributing to the emission for the targeted transitions. Emission below 5$\sigma_\mathrm{rms}$ is excluded in the integration.}
         \label{fig:moment1}
   \end{figure*}
\end{appendix}
\end{document}